\documentclass[%
reprint,
nofootinbib,
amsmath,
amssymb,
aps, 
physrev,
]{revtex4-2}
\usepackage{bm}
\usepackage{dcolumn}
\usepackage{graphicx}
\usepackage{hyperref}
\usepackage{mhchem}
\usepackage{siunitx}
\newcommand*\dif{\mathop{}\!\mathrm{d}}
\newcommand{\abs}[1]{\left|#1\right|}
\newcommand*\C{\mathop{}\!\text{core}}
\newcommand*\E{\mathop{}\!\text{edge}}

\newcommand*\R{\mathop{}\!\text{node}}

\newcommand*\NBI{\mathop{}\!\text{NBI}}

\begin{document}
\title{Optimizing External Sources for Controlled Burning Plasma in\\Tokamaks with Neural Ordinary Differential Equations}
\author{Zefang Liu}\email{Contact author: liuzefang@gatech.edu}
\author{Weston M. Stacey}
\affiliation{
Fusion Research Center, Georgia Institute of Technology, Atlanta, GA, USA
}
\begin{abstract}
Achieving controlled burning plasma in tokamaks requires precise regulation of external particle and energy sources to reach and maintain target core densities and temperatures. This work presents an inverse modeling approach using a multinodal plasma dynamics model based on neural ordinary differential equations (Neural ODEs). Given a desired time evolution of nodal quantities such as deuteron density or electron temperature, we compute the external source profiles, such as neutral beam injection (NBI) power, that drive the plasma toward the specified behavior. The approach is implemented within the NeuralPlasmaODE framework, which models multi-region, multi-timescale transport and incorporates physical mechanisms including radiation, auxiliary heating, and internodal energy exchange. By formulating the control task as an optimization problem, we use automatic differentiation through the Neural ODE solver to minimize the discrepancy between simulated and target trajectories. This framework transforms the forward simulation tool into a control-oriented model and provides a practical method for computing external source profiles in both current and future fusion devices.
\end{abstract}
\maketitle
\section{Introduction}

Controlling burning plasmas \cite{green2003iter} in tokamaks \cite{wesson2011tokamaks,stacey2012fusion} remains a central challenge in the development of fusion energy. In these high-temperature plasmas, deuterium-tritium fusion reactions produce energetic alpha particles that heat the plasma and help sustain continued fusion. While this self-heating is essential for ignition and high performance, it also introduces nonlinear feedback and potential instabilities, such as thermal runaway, if not properly controlled. To maintain stable operation, it is crucial to regulate particle densities and temperatures across different plasma regions.

Most existing models \cite{wang1997simulation,stacey1997thermal,cordey2005scaling,hill2017confinement,stacey2021nodal,liu2020one,liu2021multi,liu2022multi,liu2022thesis} simulate plasma behavior in a forward manner: external sources such as heating and fueling are specified, and the resulting plasma evolution is predicted. These models are valuable for diagnostics, transport studies, and scenario validation, but they are not designed for solving the inverse problem. In practical control tasks, the desired plasma state is specified first, and the challenge lies in determining what external source settings will achieve that state. Addressing this inverse problem is more difficult due to the nonlinear, coupled, and time-dependent nature of burning plasma dynamics.

This work proposes a solution using a physics-based multinodal transport model integrated with neural ordinary differential equations (Neural ODEs) \cite{chen2018neural,faraji2025machine}. By modeling the system as a differentiable dynamical system, we can compute gradients of a loss function with respect to time-dependent external sources. This enables direct optimization of inputs such as neutral beam injection (NBI) \cite{hemsworth2008status,hemsworth2017overview} power to achieve a target evolution of plasma quantities. The proposed framework supports control-oriented modeling and enables end-to-end optimization of external source profiles. It is implemented in the NeuralPlasmaODE \cite{liu2024application,liu2024application2} package and is applicable to both present-day experiments and future devices such as ITER \cite{aymar2002iter,holtkamp2007overview}.

\section{Methodology}

We develop our control framework based on the NeuralPlasmaODE \cite{liu2024application,liu2024application2} model, which represents tokamak plasma evolution through a set of coupled ordinary differential equations (ODEs) across multiple spatial regions. Each region, including the core, edge, scrape-off layer, and divertor, is modeled as a distinct node. Physical quantities such as particle densities and thermal energies are evolved according to particle and energy balance equations that incorporate source terms, transport processes, and loss mechanisms.

For each ion species $\sigma \in \{\ce{D}, \ce{T}, \alpha\}$, the particle balance equation in the core node is written as:
\begin{equation}
\frac{\dif n_{\sigma}^{\C}}{\dif t} = S_{\sigma, \text{ext}}^{\C} + S_{\sigma, \text{fus}}^{\C} + S_{\sigma, \text{tran}}^{\C} , \label{eqn:n-sigma-core}
\end{equation}
where $n_{\sigma}^{\C}$ is the density, $S_{\sigma, \text{ext}}^{\C}$ is the external source (such as $\NBI$), $S_{\sigma, \text{fus}}^{\C}$ represents production or consumption due to fusion, and $S_{\sigma, \text{tran}}^{\C}$ captures internodal particle transport. The energy balance equation for ion species $\sigma$ in the core is:
\begin{equation}
\frac{\dif U_{\sigma}^{\C}}{\dif t} = P_{\sigma, \text{aux}}^{\C} + P_{\sigma, \text{fus}}^{\C} + Q_{\sigma}^{\C} + P_{\sigma, \text{tran}}^{\C} ,
\end{equation}
and for electrons:
\begin{equation}
\begin{split}
\frac{\dif U_{e}^{\C}}{\dif t} & = P_{\Omega}^{\C} + P_{e, \text{aux}}^{\C} + P_{e, \text{fus}}^{\C} - P_{R}^{\C} \\
& \quad + Q_{e}^{\C} + P_{e, \text{tran}}^{\C} ,
\end{split}
\end{equation}
where $U_{\sigma}^{\C} = \tfrac{3}{2} n_{\sigma}^{\C} T_{\sigma}^{\C}$ is the thermal energy density. The right-hand side terms correspond to ohmic heating ($P_{\Omega}^{\C}$), auxiliary heating ($P_{\sigma, \text{aux}}^{\C}, P_{e, \text{aux}}^{\C}$), fusion heating ($P_{\sigma, \text{fus}}^{\C}, P_{e, \text{fus}}^{\C}$), radiation losses ($P_{R}^{\C}$), collisional energy transfer ($Q_{\sigma}^{\C}, Q_{e}^{\C}$), and energy transport ($P_{\sigma, \text{tran}}^{\C}, P_{e, \text{tran}}^{\C}$). Balance equations for other nodes are formulated similarly, as described in prior work \cite{liu2024application,liu2024application2}.

To formulate the inverse problem, we consider a given target trajectory of a quantity such as the core deuteron density $\tilde{n}_{\ce{D}}^{\C}(t)$, and aim to infer the external source $P_{\ce{D}, \NBI}^{\C}(t)$ that will reproduce it. The simplified evolution equation is:
\begin{equation}
\frac{\dif n_{\ce{D}}^{\C}(t)}{\dif t} = f\left(P_{\ce{D}, \NBI}^{\C}, D_{\ce{D}}^{\C}, t \right) ,
\end{equation}
where $D_{\ce{D}}^{\C}$ is the core deuteron diffusivity and $f(\cdot)$ includes transport, fusion, and other modeled physics as in Equation \ref{eqn:n-sigma-core}. The diffusivity can be modeled using a parametric formula previously introduced \cite{liu2024application}:
\begin{equation}
\begin{split}
\frac{D^{\R}_{\sigma}}{\SI{1}{m^2/s}} &= \alpha_H \left( \frac{B_T}{\SI{1}{T}} \right)^{\alpha_{B}} \left( \frac{n_e}{\SI{E19}{m^{-3}}} \right)^{\alpha_n} \left( \frac{T_e}{\SI{1}{keV}} \right)^{\alpha_T} \\
& \quad \cdot \left( \frac{\abs{\nabla T_e}}{\SI{1}{keV/m}} \right)^{\alpha_{\nabla T}} q(\rho)^{\alpha_q} \kappa^{\alpha_{\kappa}} \left( \frac{M}{\SI{1}{amu}} \right)^{\alpha_M} \\
& \quad \cdot \left( \frac{R}{\SI{1}{m}} \right)^{\alpha_{R}} \left( \frac{a}{\SI{1}{m}} \right)^{\alpha_{a}} ,
\end{split}
\end{equation}
where $\alpha_H, \alpha_B, \dots, \alpha_a$ are trainable parameters. These parameters have already been inferred from experimental databases such as DIII-D \cite{liu2024application,liu2024application2}, allowing this work to directly reuse the fitted diffusivity model without additional training.

To solve the inverse problem, we define a loss function measuring the deviation between the simulated and desired core deuteron density:
\begin{equation}
\begin{split}
\mathcal{L}_{n_{\ce{D}}}^{\C} &= \frac{1}{T} \int_0^T \left( n_{\ce{D}}^{\C}(t) - \tilde{n}_{\ce{D}}^{\C}(t) \right)^2 \dif t \\
& \approx \frac{1}{n_t} \sum_{t_i} \left( n_{\ce{D}}^{\C}(t_i) - \tilde{n}_{\ce{D}}^{\C}(t_i) \right)^2 , \label{eqn:loss-function}
\end{split}
\end{equation}
and compute the gradient $\partial \mathcal{L}_{n_{\ce{D}}}^{\C} / \partial P_{\ce{D}, \NBI}^{\C}$ via automatic differentiation \cite{baydin2018automatic,paszke2019pytorch}. This enables optimization of $P_{\ce{D}, \NBI}^{\C}(t)$ using standard gradient-based algorithms. The same methodology can be extended to optimize source terms for other target variables, such as the desired core deuteron temperature $\tilde{T}_{\ce{D}}^{\C}(t)$ or edge electron temperature $\tilde{T}_{e}^{\E}(t)$.

By transforming a forward simulation into an inverse design process, this approach enables computation of actionable control strategies for burning plasma systems, grounded in a physically consistent and differentiable model. This formulation bridges predictive modeling and real-time control by directly linking target plasma behaviors with the external sources needed to achieve them, offering a scalable framework for optimizing external sources in both experimental planning and reactor operation.

\section{Implementation}

\begin{figure}[h]
    \centering
    \includegraphics[width=\linewidth]{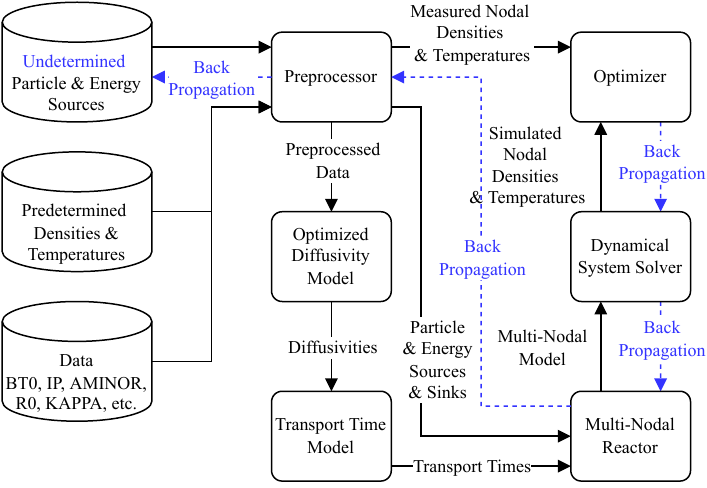}
    \caption{Workflow for optimizing particle and energy sources using NeuralPlasmaODE. Solid lines represent forward modeling, and dashed lines indicate gradients used to update sources.}
    \label{fig:workflow}
\end{figure}

We extend the NeuralPlasmaODE\footnote{\url{https://github.com/zefang-liu/NeuralPlasmaODE}} package to enable inverse source optimization for burning plasma control. The modified workflow is shown in Figure~\ref{fig:workflow}, where solid arrows indicate forward simulation of plasma evolution, and dashed arrows represent gradient computations and parameter updates. The process begins by initializing external particle and energy sources, such as neutral beam injection (NBI) \cite{hemsworth2008status,hemsworth2017overview} or auxiliary heating \cite{iter1999chapter,inoue2001iter,imai2001iter}, using empirical formulas or heuristic estimates. The model then performs forward integration of the coupled ODE system to simulate the resulting plasma state. A loss function as Equation \ref{eqn:loss-function} is evaluated by comparing selected nodal quantities, such as density or temperature, with user-defined target trajectories. Gradients of the loss with respect to the time-dependent source terms are computed using automatic differentiation through the Neural ODE solver. These gradients are used in a gradient-based optimization loop that iteratively updates the external source profiles, guiding the system toward the desired plasma behavior.

\section{Conclusion}

In this work we present a simulation-based control framework for optimizing external particle and energy sources to achieve controlled burning plasma behavior in tokamaks. By integrating multinodal plasma transport models with neural ordinary differential equations, the method enables differentiable modeling and gradient-based optimization of input profiles such as neutral beam injection. Although this paper focuses on the formulation and implementation of the approach, the framework is general and can be extended to support additional control objectives such as multi-variable trajectory tracking, real-time feedback, and actuator constraints, as well as integration with experimental data for validation and deployment.

To advance toward practical deployment, we plan to validate the proposed method using experimental data from DIII-D to assess its performance under realistic plasma conditions. Building on the current model, future efforts will apply the framework to full deuterium-tritium scenarios and focus on optimizing time-dependent source profiles for both particle and temperature control. This work supports the development of control-oriented modeling tools and contributes to predictive, model-based control strategies for ITER and other advanced magnetic confinement fusion experiments.

\bibliography{apssamp}
\end{document}